\documentclass[prc,aps,showpacs]{revtex4}
\usepackage{graphicx}
\newcommand{\be}{\begin{eqnarray}}

\newcommand{\beq}{\begin{equation}}
\newcommand{\eeq}{\end{equation}}

\newcommand{\ee}{\end{eqnarray}}
\newcommand{\ba}{\begin{array}}
\newcommand{\ea}{\end{array}}

\newcommand{\at}{\overline{10}}

\begin{document}

\title{Nonstrange and other flavor partners of the exotic
$\Theta^+$ baryon}

\author{
R.~A.~Arndt$^1$\footnote[1]{Email: arndt@reo.ntelos.net},
Ya.~I.~Azimov$^2$\footnote[2]{Email: azimov@pa1400.spb.edu},
M.~V.~Polyakov$^2$$^3$$^4$\footnote[3]{Email: Maxim.Polyakov@ulg.ac.be},
I.~I.~Strakovsky$^1$\footnote[4]{Email: igor@gwu.edu},
R.~L.~Workman$^1$\footnote[5]{Email: rworkman@gwu.edu}
\vspace*{0.1in}
}

\affiliation{
$^1$Center for Nuclear Studies, Department of Physics,
    The George Washington University, Washington, D.C.
    20052, USA \\
$^2$Petersburg Nuclear Physics Institute, Gatchina,
    St.~Petersburg 188300, Russia \\
$^3$Universit\'e  de Li\`ege au Sart Tilman,
                   B-4000 Li\`ege 1 Belgium \\
$^4$Institute for Theoretical Physics~II, Ruhr University,
    44780 Bochum, Germany \\
}

\begin{abstract}

Given presently known empirical information about the
exotic $\Theta^+$ baryon, we analyze possible properties
of its $SU(3)_F$-partners, paying special attention to
the nonstrange member of the antidecuplet $N^{\ast}$.
The modified PWA analysis presents two candidate masses,
1680~MeV and 1730~MeV. In both cases the $N^{\ast}$ should
be highly inelastic. The theoretical analysis, based on
the soliton picture and assumption of $\Gamma_{\Theta^+}
<5$~MeV, shows that most probably $\Gamma_{N^{\ast}}<30$~
MeV.  Similar analysis for $\Xi_{3/2}$ predicts its width
to be not more than about 10~MeV. Our results suggest
several directions for experimental studies that may
clarify properties of the antidecuplet baryons, and
structure of their mixing with other baryons.

\end{abstract}


\maketitle

\section{Introduction}
\label{sec:intro}

The observation of narrow peaks in the invariant mass
spectra of $nK^+$~\cite{leps,clas1,clas2,elsa,clas3} and
$pK_S$~\cite{itep1,itep2,herm,zeus,na49} events, verified
by independent groups and laboratories worldwide, has
solidified the evidence for an exotic baryon $\Theta^+$
with strangeness $+1$, a mass of about 1540~MeV, and a
narrow width.  The existence of such a particle implies
a whole new family of $SU(3)_F$-partners, beyond the
familiar octets and decuplets.  This state was predicted
(both the mass and the narrow width)~\cite{dpp} on the
basis of a chiral soliton approach to hadron dynamics.
In this approach it should be a member of a flavor
antidecuplet with $J^P = 1/2^+$.

The antidecuplet had emerged even earlier in various
versions of the soliton approach to 3-flavor QCD (for a
brief review of the history and earlier references see,
\textit{e.g.}, \cite{dpp,kopwal,pr1}). However, the
expected masses of its members had always been rather
uncertain, at least up to \mbox{$\sim 100$~MeV}. Until
Ref.~\cite{dpp}, the question of width, also essential
for experimental searches, had not been addressed at
all.  To make the $\Theta^+$-mass prediction more
definite, it was suggested in~\cite{dpp} to identify
the non-strange member of the antidecuplet with the
$N(1710)$, the only nucleon state listed in the
PDG-tables~\cite{PDG02} having $J^P = 1/2^+$ and being
in the expected mass interval.  Clearly, this assignment
resulted in a good agreement with later experimental
findings for the $\Theta^+$ mass.  Given the present
more detailed knowledge of
$\Theta^+$-properties, and with higher-statistics
experiments under preparation, it is important to
reconsider the antidecuplet nature of the $N(1710)$.

Let us first summarize our present knowledge of the
$\Theta^+$-width.  The theoretical prediction,
simultaneous with its mass, was $\Gamma_{\Theta^+}<
15$~MeV~\cite{dpp}, unexpectedly narrow for strong
decays.  Existing measurements, instead of determining
$\Gamma_{\Theta^+}$,  have only shown it to be smaller
than experimental resolution (see Table~\ref{tbl1}).
Most experimental publications have given an upper
bound of about $\sim 20$~MeV.  Xenon bubble chamber
data, corresponding in essence to the charge exchange
reaction $K^+n\to K^0p$, have provided the slightly
lower bound of 9~MeV~\cite{itep1}.

Less direct determinations~\cite{nuss}, using previously
measured $K^+d$ total cross sections, have led to a
stronger limitation $\Gamma_{\Theta^+}<6$~MeV.  A
similar bound, $<5$~MeV, was obtained in Ref.~\cite{hk}
within a more elaborated theoretical description.  The
partial-wave analysis (PWA) of available $KN$ (elastic
and charge exchange) scattering data similarly claims
to exclude widths above 1--2~MeV~\cite{asw}.  A more
detailed reexamination of the approach in 
Ref.~\cite{nuss} provides nearly the same result of 
$\Gamma_{\Theta^+}< 1.5$~MeV~\cite{asw1}.  A similar 
method applied to the Xenon data~\cite{itep1} has allowed 
even the tentative claim of a lower limit $\Gamma
_{\Theta^+} = 0.9\pm0.3$~MeV~\cite{cahntr} (with 
additional assumptions and an unknown systematic 
uncertainty).  One should emphasize, however, that all 
of these indirect treatments assume the existence of a 
$\Theta^+$, which they can not confirm.  Moreover, 
they are based mainly on rather old data, which may be 
shifted by the next generation of higher precision 
measurements.  Nevertheless, we should take these 
results into account when discussing the $\Theta^+$ as 
given by the present data.

Evidently, all of the above estimates for $\Gamma
_{\Theta^+}$ are in sharp contrast with the width $\sim
100$~MeV ascribed~\cite{PDG02} to the $N(1710)$,
initially considered to be a unitary partner of the
$\Theta^+$~\cite{dpp}.  Of course, members of the same
unitary multiplet can have different widths, but in the
absence of a special reason (say, mixing with members
of another multiplet) it would be more natural for them
to have comparable widths.

\begin{figure}[th]
\centerline{
\includegraphics[height=0.5\textwidth, angle=90]{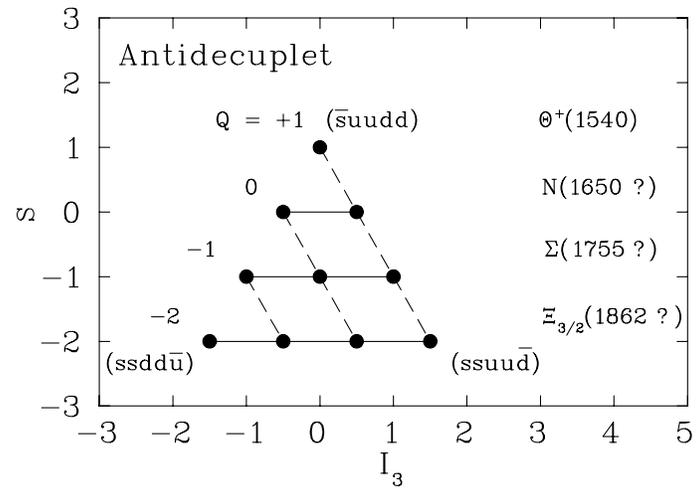}}
\caption{Tentative unitary anti-decuplet with $\Theta^+$.
         Isotopic multiplet (constant values of the charge)
         shown by solid (dashed) lines.
\label{fig:g1}}
\end{figure}

Additional information related to the assignment of
unitary partners is due to a recent experimental
result~\cite{xi} giving evidence for one further
explicitly exotic particle $\Xi^{--}_{3/2}$, with a
mass $1862\pm2$~MeV and width $< 18$~MeV (\textit{i.e.}
less than resolution). Such a particle had been expected
to exist as a member of the same antidecuplet containing
the $\Theta^+$, but its mass was predicted to be about
2070~MeV~\cite{dpp}, essentially different from the
experimental value.  This has posed similar problems
for the masses of other unitary partners of the
$\Theta^+$, nucleon-like and $\Sigma$-like.  The
supposed antidecuplet looks today as shown on 
Fig.~\ref{fig:g1}, with $\Sigma$- and $N$-masses 
determined by the Gell-Mann--Okubo rule.

The state $N(1710)$, though listed in the PDG Baryon
Summary Table~\cite{PDG02} as a 3-star resonance, is
not seen in a recent analysis of pion-nucleon
elastic scattering data~\cite{aswp}.  Studies which
have claimed to see this state have given widely
varying estimates of its mass and width (from $\sim
1680$~MeV to $\sim 1740$~MeV for the mass and from $\sim
90$~MeV to $\sim 500$~MeV for the width). Branching ratios
have also been given with large uncertainties (10--20\%
for $N\pi$, 40--90\% for $N\pi\pi$, and so on), apart
from one which has been presented with greater
precision ($6\pm 1\%$ for $N\eta$).

Of course the non-observation of a broad $N(1710)$
state in pion-nucleon elastic analyses could be due to
a very small $\pi N$ branching ratio.  Standard
procedures used in partial-wave analysis (PWA) may also
miss narrow resonances with $\Gamma< 30$~MeV (a similar
situation below inelastic thresholds has been discussed
in~\cite{aasw}).  Therefore, the true unitary partner
of the $\Theta^+$ (if it is different from  $N(1710)$
and sufficiently narrow) could have eluded detection.

Here we reconsider the identity of $N^{\ast}$, the
nucleon-like partner of $\Theta^+$, and investigate the
possible existence and properties of narrow non-strange
state(s) near 1700~MeV.  We first consider modifications 
of a PWA with narrow resonances, and apply the results 
to $\pi N$ elastic scattering at $W\sim 1700$~MeV 
(Section~\ref{sec:pwa}).  Section~\ref{sec:theor} presents 
a discussion of possible properties of the $N^{\ast}$ in 
the soliton picture with baryon mixing and for a small 
$\Theta^+$ width.  Some expected properties of the 
$\Xi_{3/2}$ are also considered.  Our results are briefly 
discussed and conclusions formulated in 
Section~\ref{sec:disc}.

\section{Narrow resonances in partial-wave analyses}
\label{sec:pwa}

We have emphasized earlier~\cite{aasw} that standard
methods of PWA are insensitive to very narrow resonances.
Therefore, a modified approach is required to search for
the presence of a narrow resonance with particular
values of mass and width~\cite{aasw} (see also~
\cite{asw}).  We consider the situation in more detail,
separately for elastic and inelastic cases.

\subsection{Elastic case}
\label{sec:el}

Interaction in the elastic case may transform a state $a$
only to a similar state $a'$ (changing, for example,
particle momenta without changing particle identity).
One can then choose physical states, so as to diagonalize
the $S$ matrix, (\textit{e.g.}, for the $\pi N$ scattering,
take states with definite values of energy, isospin,
parity, and angular momentum), and have only diagonal
transitions $a\to a$ with $S$ matrix elements
\begin{eqnarray}
\langle a|S|a\rangle = e^{2i\delta_a}\,.
\label{1}\end{eqnarray}
Standard methods employ some parametrization of the
interaction phase $\delta_a$, fitting these parameters
to describe experimental data.  Instead, we will
split the phase as
\begin{eqnarray}
\delta_a = \delta_a^B + \delta_R\,.
\label{2}\end{eqnarray}
The background part $\exp(2i\delta_a^B)$ may be
parametrized as before, while the resonance part has the
canonical Breit-Wigner form
\begin{eqnarray}
e^{2i\delta_R} = \frac{M_R - W + i\Gamma_R/2}{M_R - W -
i\Gamma_R/2}\,.
\label{3}\end{eqnarray}
If refitting (over the whole database) with some fixed
values of $M_R$ and $\Gamma_R$ provides a worse
description (higher $\chi^2$) than without the resonance,
then a resonance $R$ with the corresponding mass and
width is unsupported.  If the new description is better
(has lower $\chi^2$), then the resonance \textit{may}
exist.

At first sight, we have increased the number of
parameters and, therefore, should always have a better
description.  This is not necessarily so, due to the
specific form used to introduce these two additional
parameters, and the fixed values assigned to them.
Moreover, we have demonstrated in Ref.~\cite{aasw} that
a better description (lower $\chi^2$) may result for
various reasons not associated with the presence of a
resonance.  Nevertheless, searching for a better
description allows us to restrict the region in
$(M_R,~\Gamma_R)$ space, where a resonance may be
assumed.  This is the approach that was used earlier
as a basis for numerical procedures which restricted
admissible widths of light $\pi N$ resonances~\cite{aasw} 
and of the $\Theta^+$~\cite{asw}.

\subsection{Inelastic case}
\label{sec:inel}

For energies near $W\sim 1700$~MeV, we may be sensitive
to thresholds for the production of additional (or
different) mesons, and one should take the inelasticity
into account even when investigating purely elastic
$\pi N$ scattering.

Because of inelastic transitions, nondiagonal $S$
matrix elements, generally, do not vanish.  However,
the $S$ matrix is a unitary operator and, hence, can be
diagonalized by a unitary transformation:
\begin{eqnarray}
S = US^{(0)}U^{-1}\,,
\label{4}\end{eqnarray}
with $S^{(0)}$ having a diagonalized form and
\begin{eqnarray}
\sum_{n\geq0}|U_{an}|^2 = \sum_a|U_{an}|^2 = 1\,,
\label{unit}\end{eqnarray}
due to unitarity.  One can present an elastic amplitude
(for example, a partial-wave amplitude) as
\begin{eqnarray}
\langle a|S|a\rangle = \sum_{n\geq 0}|U_{an}|^2 \,
e^{2i\delta_n}\,,
\label{s_mat}\end{eqnarray}
where the unitarity of $U$ has been used.  The form
(\ref{s_mat}) implies the well-known relation
\begin{eqnarray}
|\langle a|S|a\rangle|\leq\sum_{n\geq 0}|U_{an}|^2
= 1\,.
\label{7}\end{eqnarray}

If the resonance candidate $R$ is narrow enough to
avoid overlap with other resonances, the explicit
resonance behavior can be inserted in only one of
$\delta_n$, say, to $\delta_0$:
\begin{eqnarray}
\delta_0 = \delta_0^B + \delta_R\,
\label{8}\end{eqnarray}
(compare to Eq.~(\ref{2})).
Then,
\begin{eqnarray}
\langle a|S|a\rangle = |U_{a0}|^2 \,e^{2i\delta_0^B}\,
e^{2i\delta_R} + \sum_{n\geq 1}|U_{an}|^2 \,e
^{2i\delta_n}\,.
\label{inel}\end{eqnarray}
All quantities on the right hand side of Eq.$\,$
(\ref{inel}) depend on energy, but only $\delta_R$
has a sharp energy dependence near the resonance.

It is easy to see that
\begin{eqnarray}
|U_{a0}|^2|_{W = M_R} = r_a
\label{10}\end{eqnarray}
is the branching ratio for a particular decay mode
$R\to a$.  Then, Eq.~(\ref{unit}) provides the
expected relation
\begin{eqnarray}
\sum_a r_a = 1\,.
\label{11}\end{eqnarray}
Now, we can rewrite Eq.~(\ref{inel}) as
\begin{eqnarray}
\langle a|S|a\rangle = r_a\,A(W)\,e^{2i\delta_R} +
(1 - r_a)B(W)\,,
\label{12}\end{eqnarray}
where
\begin{eqnarray}
r_a\,|A(W)| + (1 - r_a)|B(W)|\leq 1\,,~~~~|A(M_R)|
= 1\,.
\label{13}\end{eqnarray}
The expressions (\ref{12}) and (\ref{13}) can be
used to construct a parametrization and numerical
procedure, similar to one described in Ref.$\,$
\cite{aswp}, to test for the existence of a possible
resonance $R$.

Note that, in the inelastic case, description of the
resonance contribution to the elastic amplitude
\mbox{$a\to a$} contains three parameters ($M_R,~
\Gamma_R,~r_a$), instead of two in the elastic
case.  However, far from the resonance, at $|W -
M_R|\gg\Gamma_R$, the contribution takes the form
$\propto\Gamma^a/(M_R - W)$, sensitive only to the
partial decay width
\begin{eqnarray}
\Gamma^a = r_a\,\Gamma_R\,.
\label{14}\end{eqnarray}
This is trivially true for a purely elastic
resonance, for which partial and total widths
coincide.

\subsection{Fitting the data}
\label{sec:fit}

Nucleon-like states may be revealed in various
processes, with various initial and final states.
But most convincing are their manifestations in
$\pi N$ elastic scattering.  As a result, we
consider here only elastic (and charge exchange)
data.

\begin{figure}[th]
\centerline{
\includegraphics[height=0.3\textwidth, angle=90]{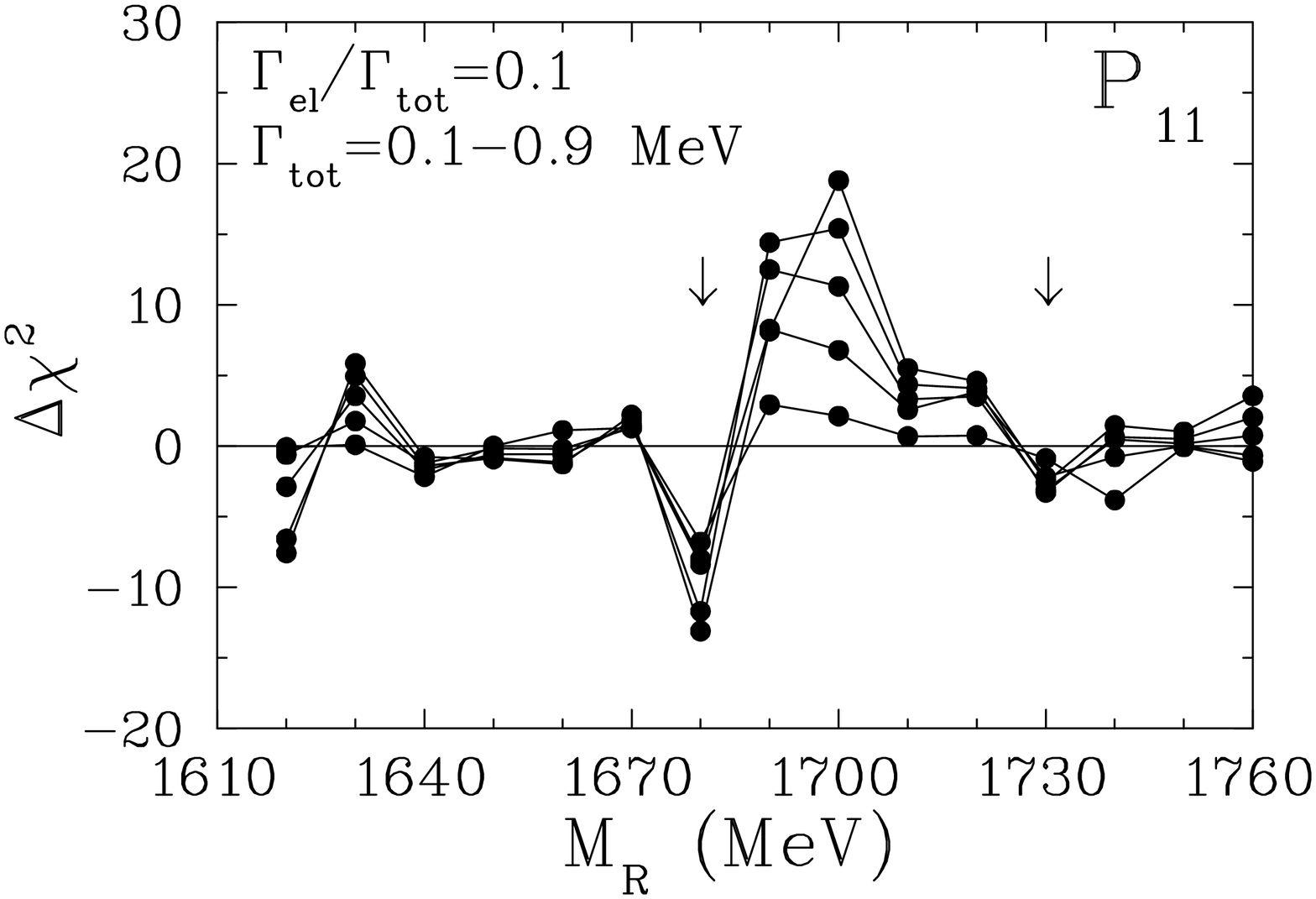}\hfill
\includegraphics[height=0.3\textwidth, angle=90]{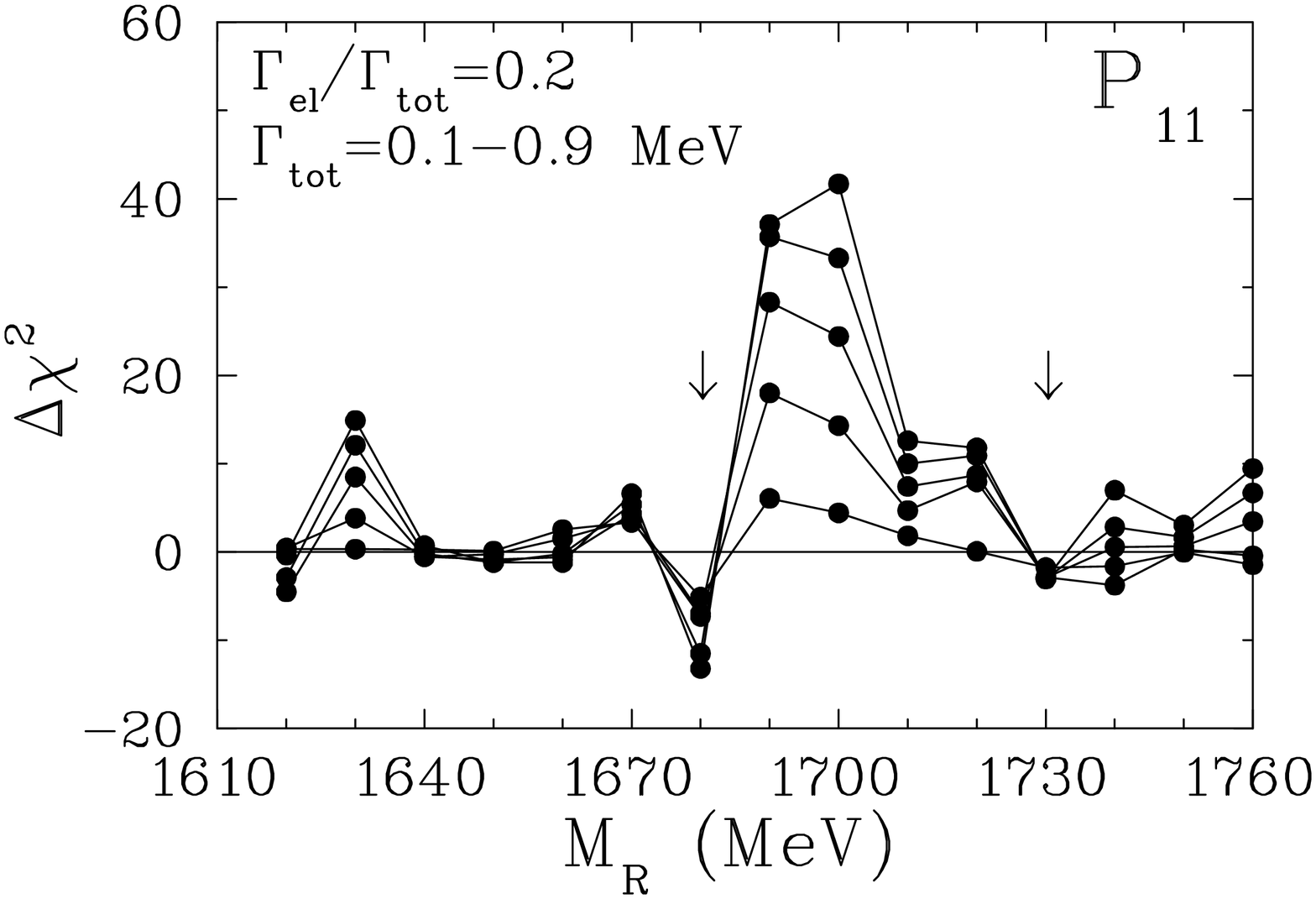}\hfill
\includegraphics[height=0.3\textwidth, angle=90]{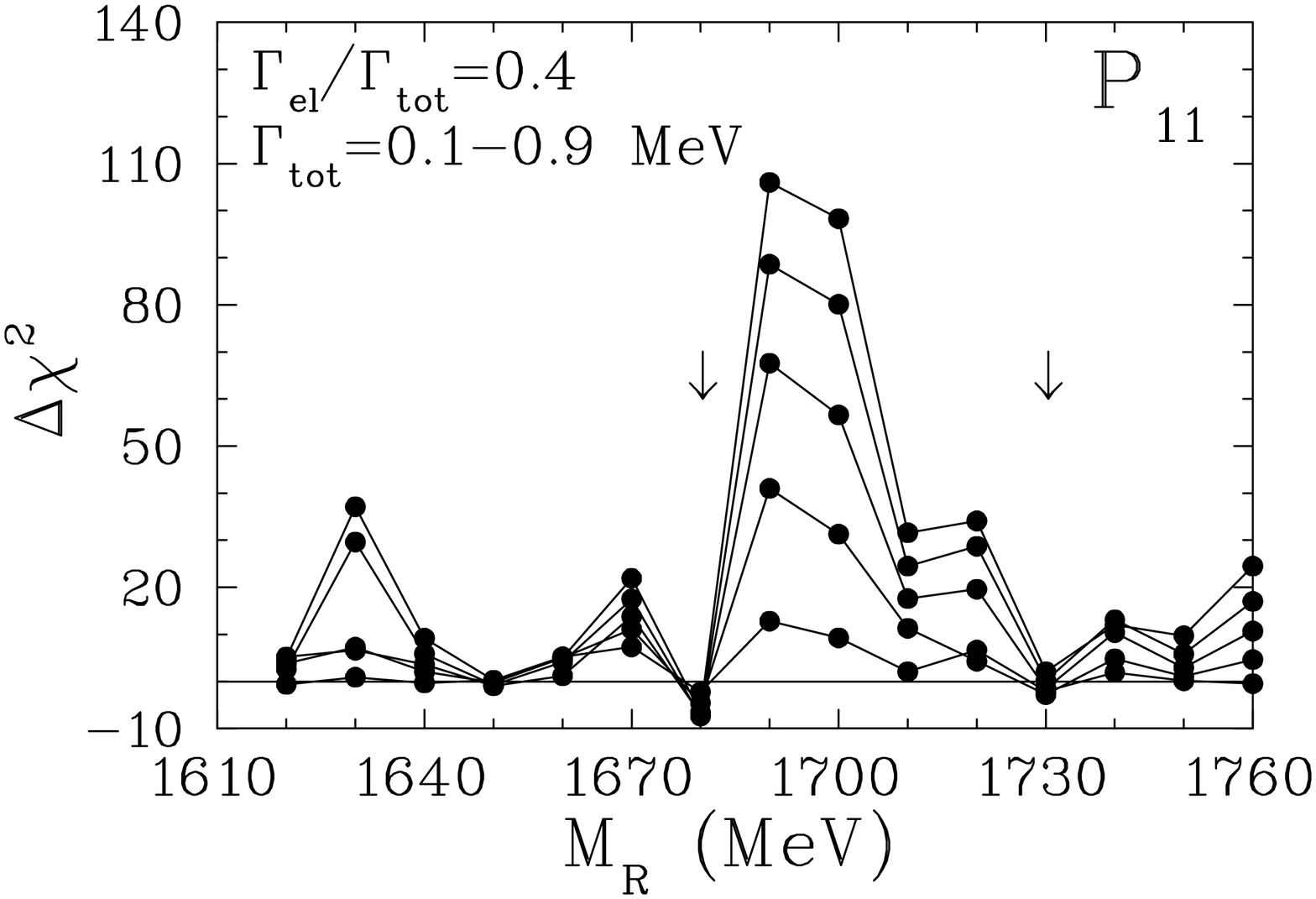}}
\centerline{
\includegraphics[height=0.3\textwidth, angle=90]{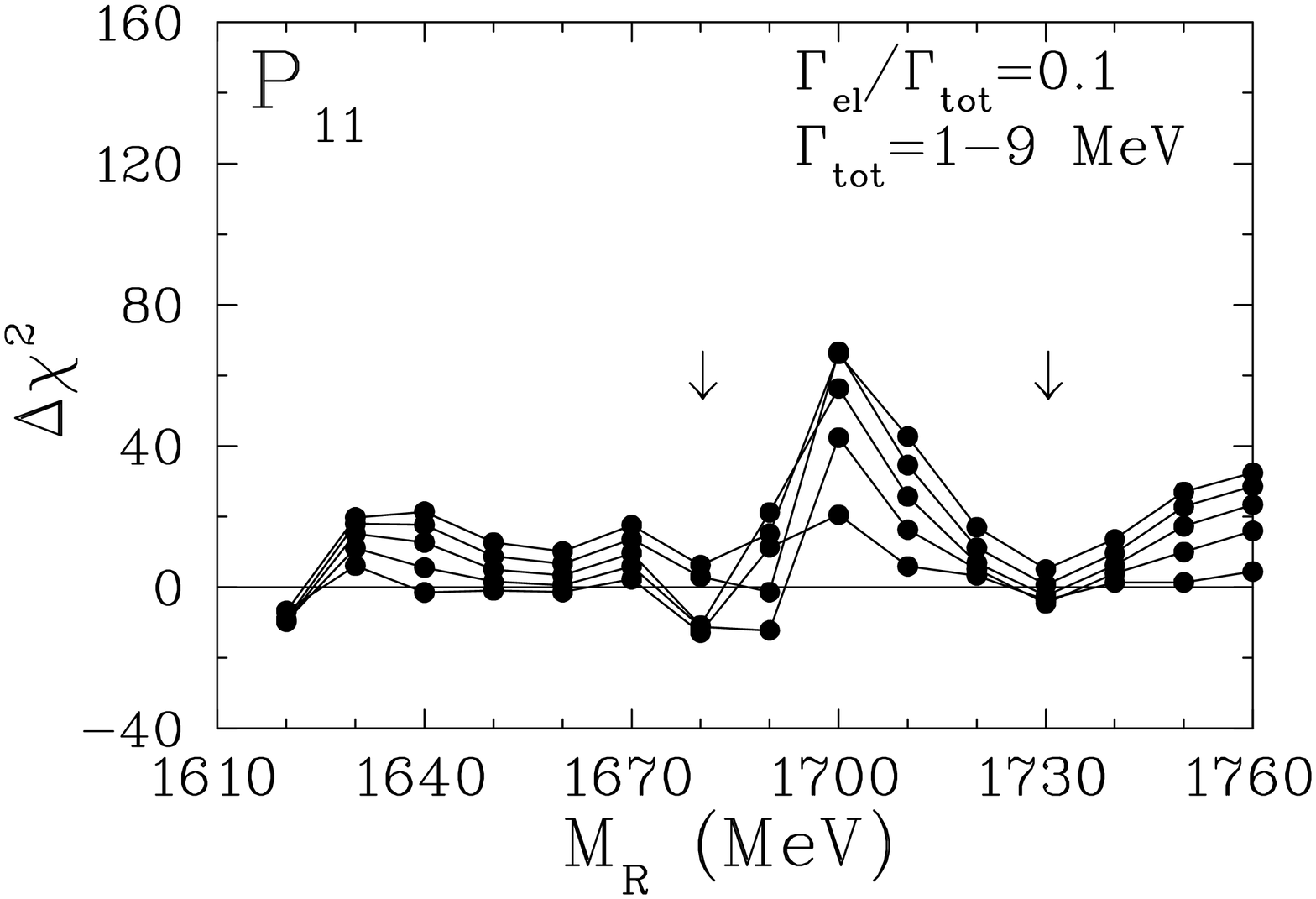}\hfill
\includegraphics[height=0.3\textwidth, angle=90]{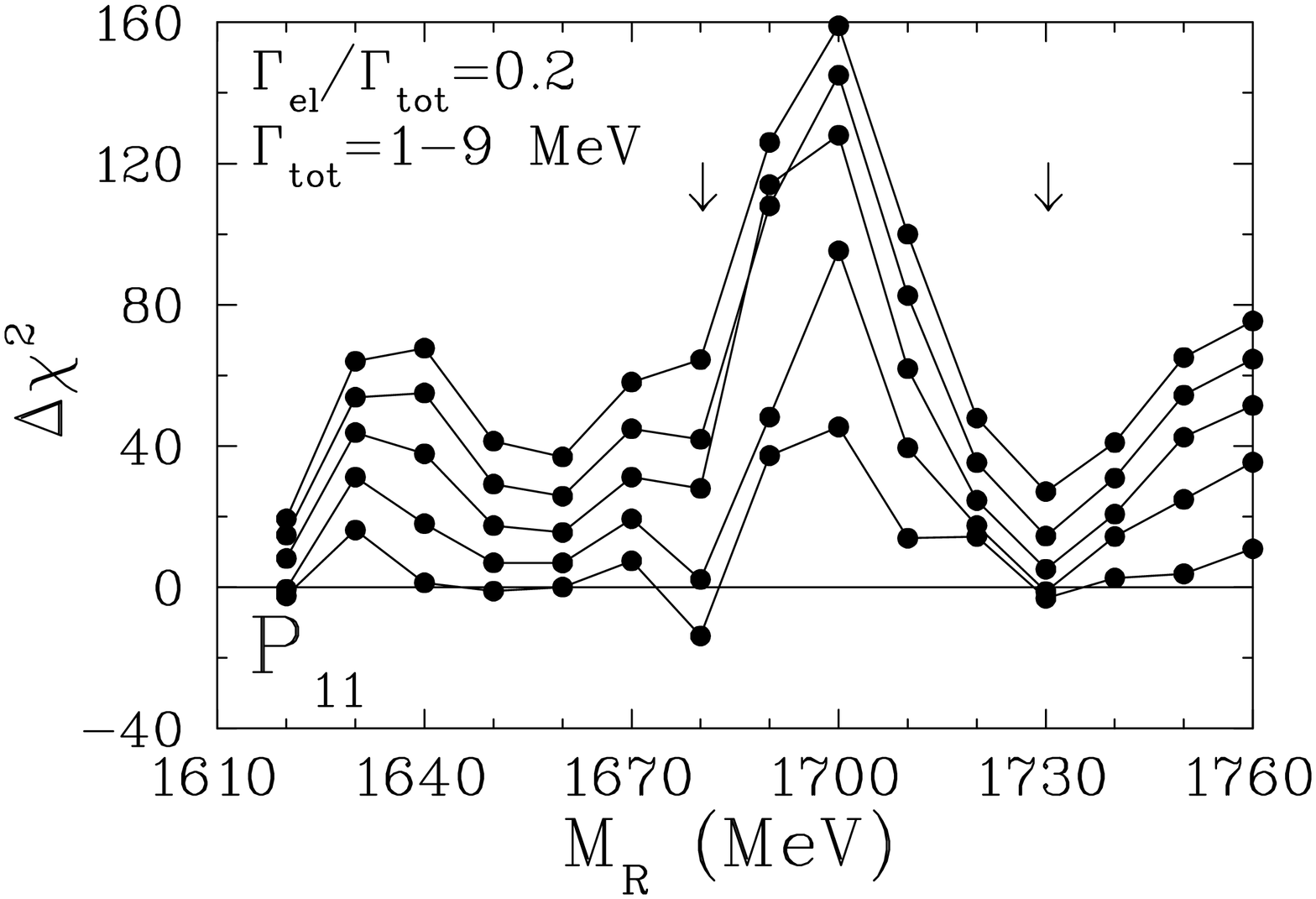}\hfill
\includegraphics[height=0.3\textwidth, angle=90]{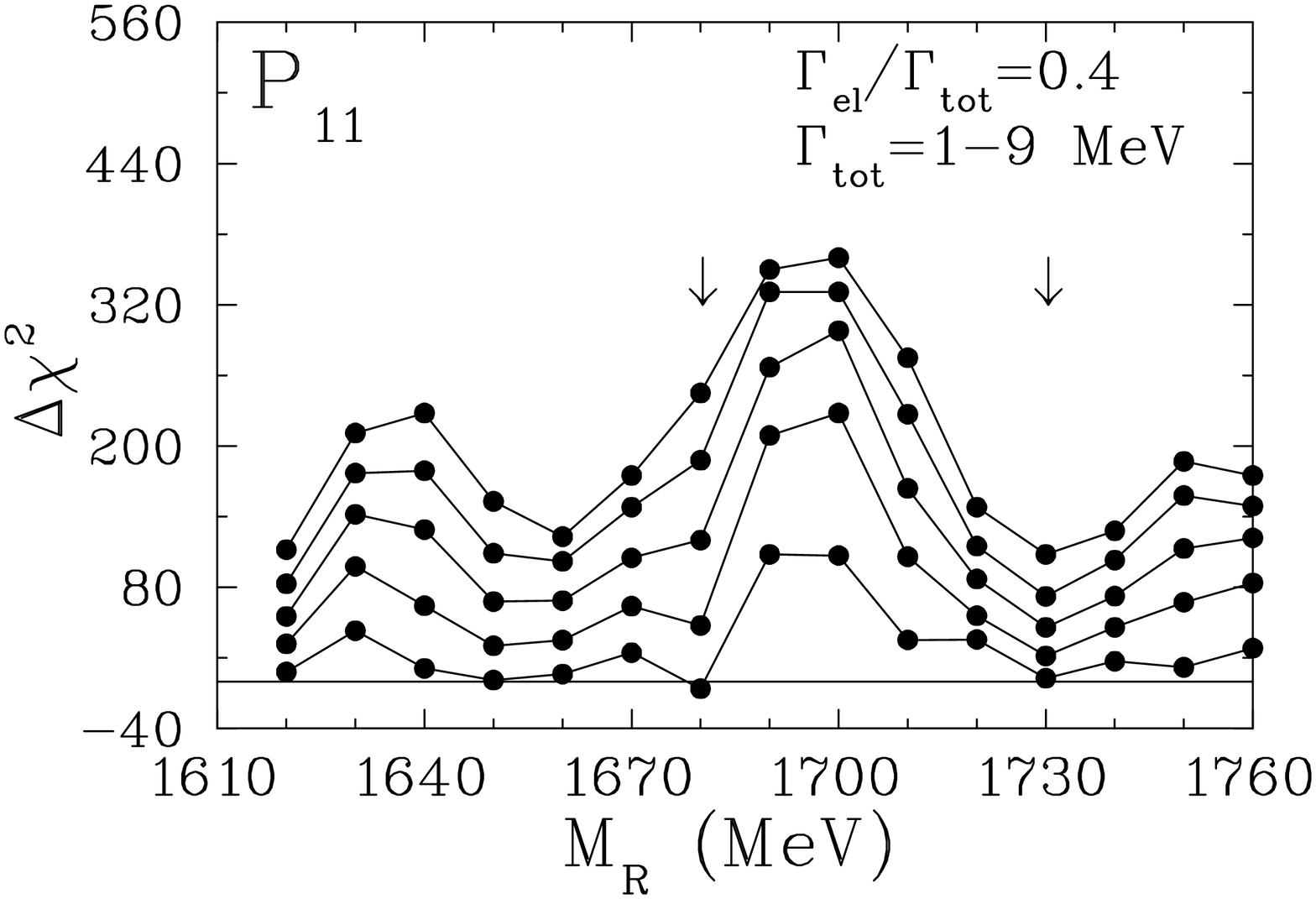}}
\caption{Change of overall $\chi^2$ due to insertion of a
         resonance into $P_{11}$ for $M_R$ = 1660 -- 1760~MeV
         with $\Gamma_{tot}$ = 0.1, 0.3, 0.5, 0.7, and
         0.9~MeV (top panel) and 1, 3, 5, 7, and 9~MeV
         (bottom panel), and $\Gamma_{el}/\Gamma_{tot}$
         = 0.1 (left column), 0.2 (midle column), and
         0.4 (right column) using $\pi N$ PWA~
         \protect\cite{aswp}.  The curves are given to
         guide the eye.  Vertical arrows indicate M$_R$ =
         1680 and 1730~MeV. \label{fig:g2}}
\end{figure}

We begin by considering the $\pi N$ partial wave
$P_{11}$, as this amplitude is associated with
resonances having $J^P = 1/2^+$.  The character
of $\chi^2$-changes, $\Delta\chi^2$, after inserting
a narrow resonance with a range of masses, widths,
and branching fractions is illustrated in Fig.$\,$
\ref{fig:g2}.  The resonance mass has been allowed
to vary from 1620 to 1760~MeV in 10~MeV steps.  For
the total width, we have used five values in the
intervals 0.1$-$0.9~MeV (step 0.2~MeV) and 1$-$9~
MeV (step 2~MeV).  For easier tracing, we have
connected points having consecutive values of the
mass and identical values for the other parameters.

Negative values of $\Delta\chi^2$ emerge most
readily near $M_R = 1680$~MeV and 1730~MeV.  We see
that $\Delta\chi^2$ becomes negative only for
$\Gamma_{el} = (\Gamma_{el}/\Gamma_{tot})\cdot
\Gamma_{tot}$ within the bounds
\be
\Gamma_{el}\leq 0.5~[0.3]~{\rm MeV}
\label{15}
\ee
for $M_R = 1680~[1730]$~MeV.  The available data
can not reliably discriminate values of $\Gamma_{el}$
below these bounds.  Neither can they discriminate
the particular values of $\Gamma_{tot}$.  Note that,
for higher values of $\Gamma_{tot}$, such states
could presumably be seen in a standard PWA; however
with the above restrictions for $\Gamma_{el}$ these
resonances would be extremely inelastic and have
little effect on the elastic scattering process.
Thus, for $J^P = 1/2^+$, we see two possible mass
values for the nucleon-type resonance state(s), both
having rather small elastic ({\textit i.e.} $\pi N$)
partial widths.

It was demonstrated, however, in Ref.~\cite{aasw}
that $\Delta\chi^2< 0$ does not necessarily mean
the real existence of a resonance.  The ``resonance"
may be only an effective mechanism to introduce
corrections, \textit{e.g.}, in the presence of
unaccounted (or badly accounted) for singularities
(say, thresholds), or insufficient and/or poor
quality of data.

\begin{figure}[th]
\centerline{
\includegraphics[height=0.3\textwidth, angle=90]{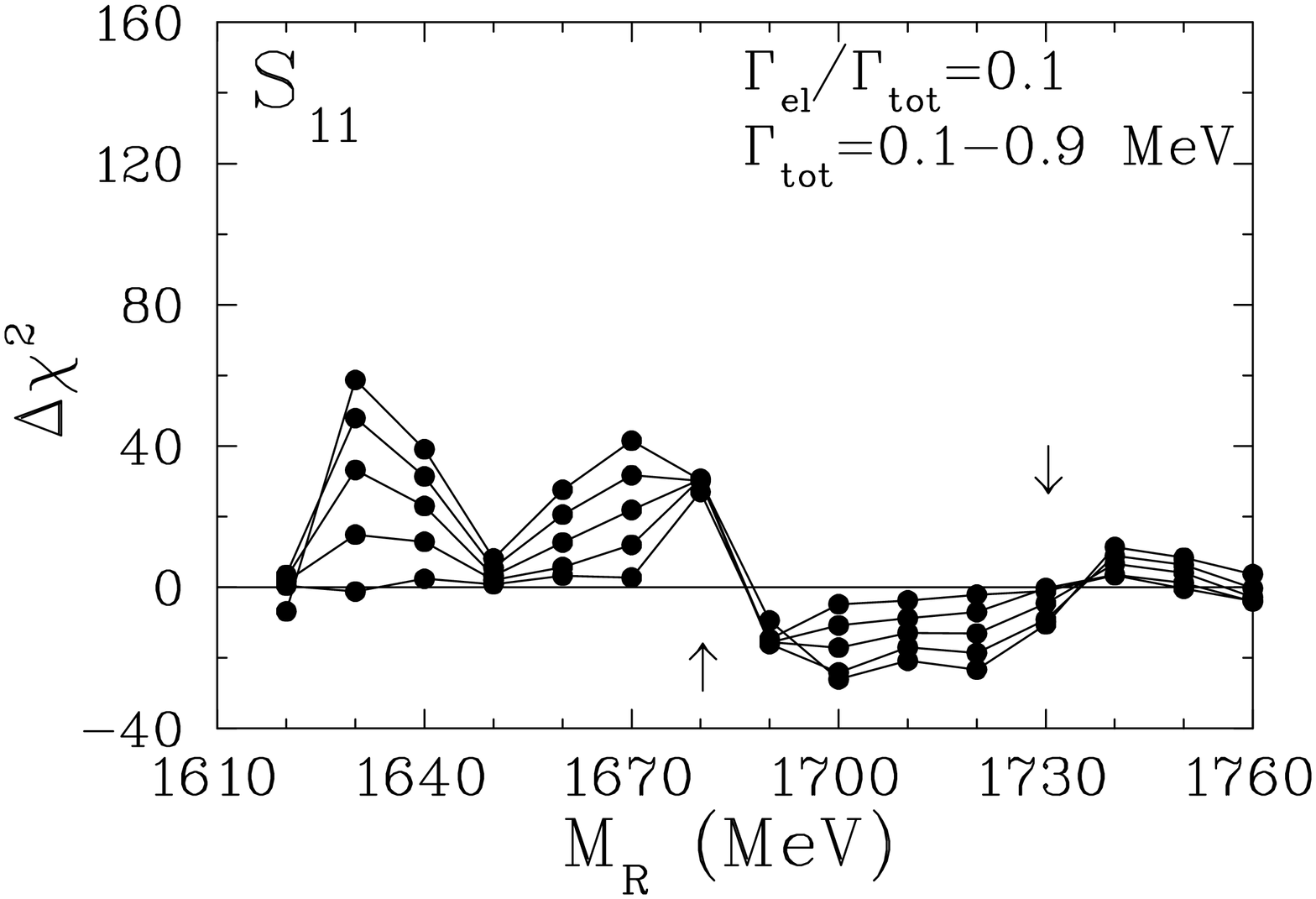}\hfill
\includegraphics[height=0.3\textwidth, angle=90]{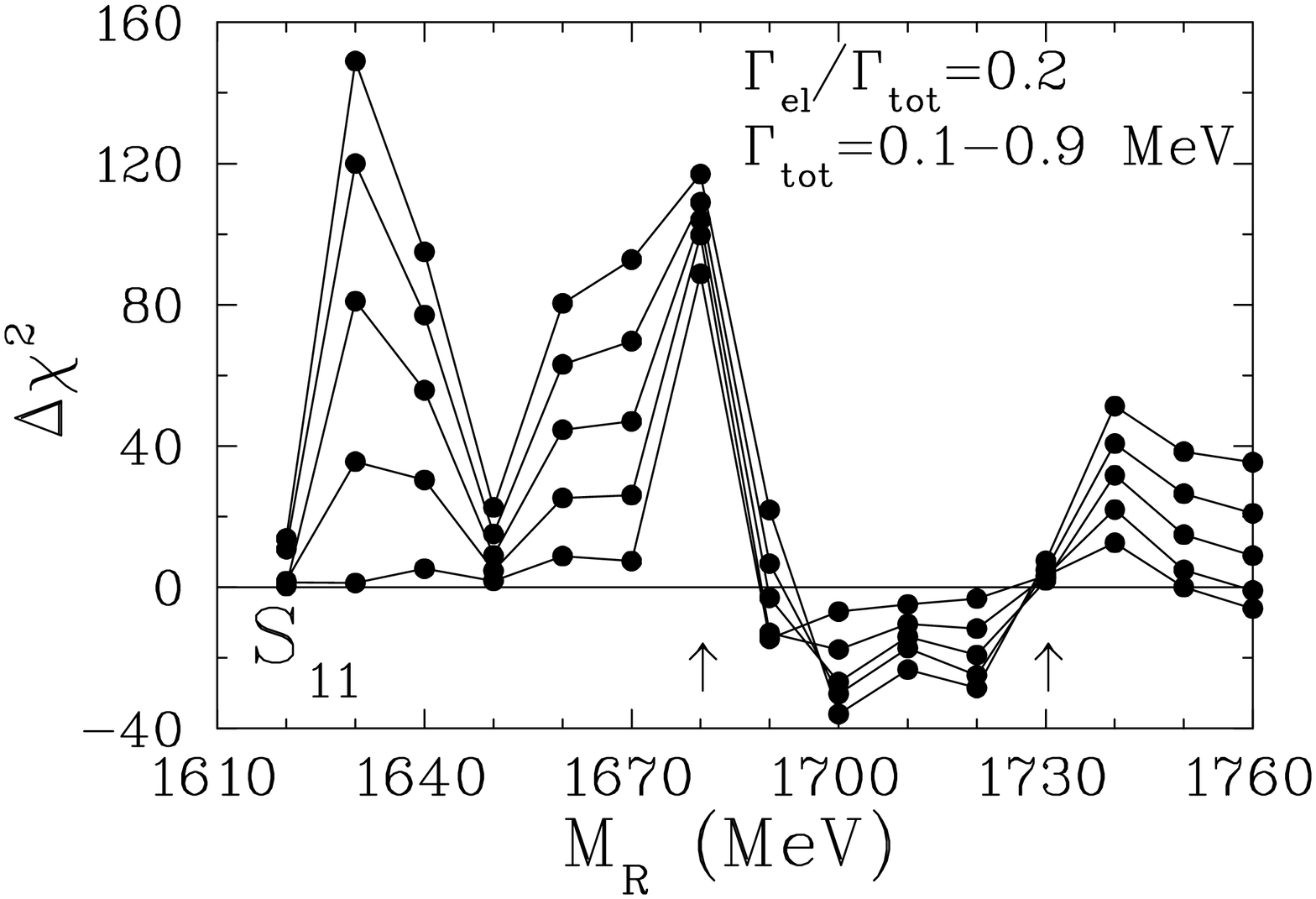}}
\centerline{
\includegraphics[height=0.3\textwidth, angle=90]{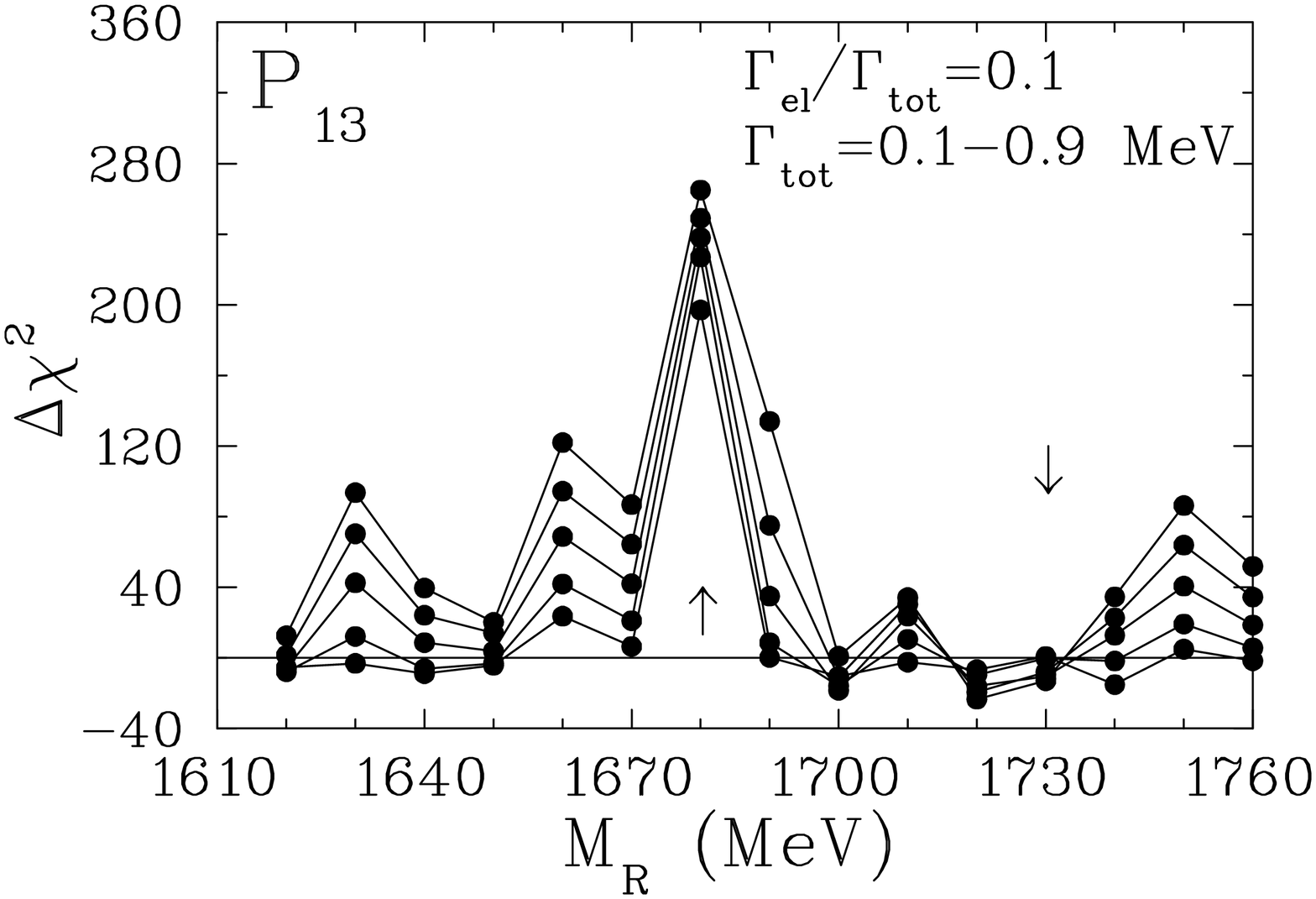}\hfill
\includegraphics[height=0.3\textwidth, angle=90]{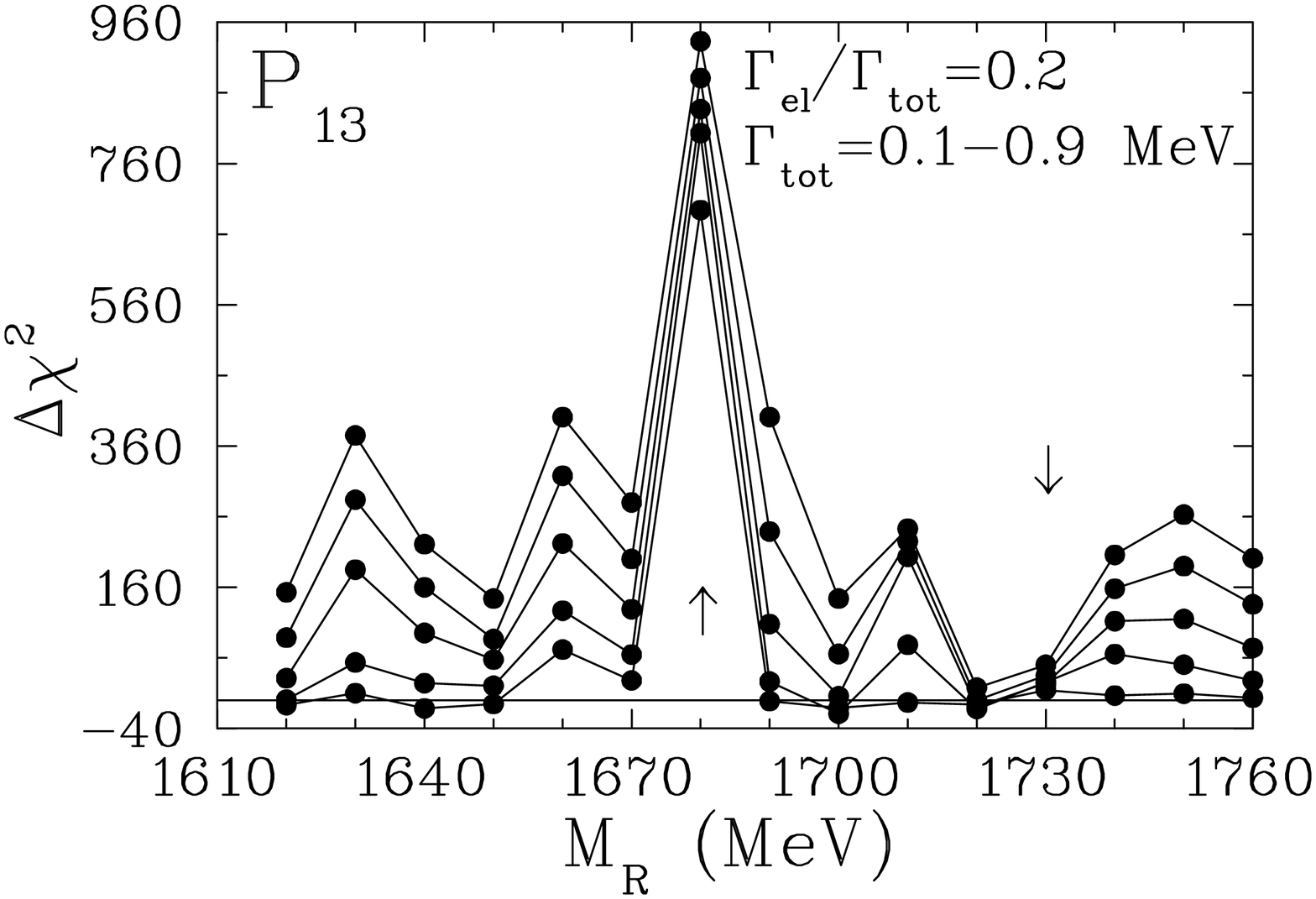}}
\caption{Change of overall $\chi^2$ due to insertion of
         resonances into $S_{11}$ (top panel) and $P_{13}$
         (bottom panel) for $M_R$ = 1660 -- 1760~MeV
         with $\Gamma_{tot}$ = 0.1, 0.3, 0.5, 0.7, and
         0.9~MeV and $\Gamma_{el}/\Gamma_{tot}$ = 0.1
         (left column) and 0.2 (right column).  Notation
         as in Fig.~\protect\ref{fig:g2}. \label{fig:g3}}
\end{figure}

A true resonance should provide an effect only when
being inserted into a particular partial amplitude,
while non-resonant sources may show sensitivity in
various partial-wave amplitudes.
To check this possibility, we have repeated the
insertion-refitting procedure for the partial-wave
amplitudes $S_{11}$ and $P_{13}$, having the
$J^P$ quantum numbers $1/2^-$ and $3/2^+$,
respectively.  Using the same values as before for
the mass, width, and branching ratio of the assumed
resonance, produces the results for $S_{11}$ and
$P_{13}$ illustrated in Fig.~\ref{fig:g3}.  No
effect emerges at $M_R$ = 1680~MeV, enhancing the
expectation of a true $P_{11}$ resonance effect at
this mass.

In the 1700--1740~MeV region, variations for $S_{11}$
and $P_{13}$ show shallow dips, somewhat similar to
one for $P_{11}$.  This may cast doubt on the
existence of a true narrow resonance in this interval.
However, the dips for $S_{11}$ and $P_{13}$ are
qualitatively different than for $P_{11}$; as a
result, we do not consider a narrow resonance with
$M_R = 1730$~MeV to be excluded.

The above dips in $\Delta\chi^2$ could be induced by
the nearby thresholds, $N\omega$ with $W_{th}\approx
1720$~MeV, and $N\rho$ with $W_{th}\approx 1710$~MeV
(note, however, its larger distance from the physical
region due to the width of the $\rho$).  It is
interesting that $M_R = 1680$~MeV appears also near
the $K\Sigma$ threshold with $W_{th}\approx 1685$~MeV.
There are many-particle thresholds as well, but their
contributions are expected to be less important than
two-particle ones, due to much smaller phase-space
near the thresholds, and we do not consider them here.
None of these thresholds have been accounted for in
the PWA parametrizations of $\pi N$ data.

Concluding this section, we emphasize that our
results suggest two possible masses for a narrow
nucleon-like resonance(s) having $J^P = 1/2^+$ and a
mass near 1700~MeV.  One of these, near 1680~MeV,
looks more promising.  Though our approach can give 
candidate values of mass and width for narrow 
resonance(s), it does not prove the existence of a 
resonance.  Therefore, all our candidates need 
further direct and detailed experimental checks.

\section{Theoretical analysis}
\label{sec:theor}

Let us discuss the above results as compared to
expected properties of the antidecuplet members in
the soliton picture.

The antidecuplet mass differences (say, between
$\Theta^+$ and $N^{\ast}$, its non-strange partner),
based on this picture and presented in Ref.$\,$
\cite{dpp}, is about 180~MeV.  Using the measured
value $M_{\Theta} = 1540$~MeV, we should obtain
$M_{N^\ast} = 1720$~MeV, which is close to the
heavier candidate mass, 1730~MeV, of the preceding
Section.

However, the soliton calculation of this mass
difference requires some assumptions.  In particular,
it depends on the value of the $\sigma$-term, which
is the subject of controversy.  Its value, taken
according to the latest data analysis~\cite{pasw},
leads to an antidecuplet mass difference of about
110~MeV \cite{dp}.

Moreover, today one is able to use another, more
phenomenological approach.  If the states
$\Xi_{3/2}$~\cite{xi} and ${\Theta}$ are indeed
members of the same antidecuplet, then, according
to the Gell-Mann-Okubo rule, the mass difference
of any two neighboring isospin multiplets in the
antidecuplet should be constant and equal
$$(M_{\Xi_{3/2}} - M_{\Theta})/3\approx 107~{\rm
MeV.}$$  This gives $M_{N^\ast}\approx 1650$~MeV,
near but lower than our lighter candidate mass,
1680~MeV.  Due to $SU(3)_F$-violating mixing with
lower-lying nucleon-like octet states, $M_{N^\ast}$
may shift upward, and reach about 1680~MeV~
\cite{dp}.  Mixing with higher-lying nucleon-like
members of exotic 27- and 35-plets may also play a
role (see Refs.~\cite{pr1,kopwal,ekp}). Note that the
effects of multiplet mixing on masses is parametrically
of order ${\cal O}(m_s^2)$ which is beyond
approximations used in Ref.~\cite{dpp}.

In discussing the expected decays for the antidecuplet 
candidates, we mainly follow Ref.~\cite{dpp}.  Though 
expressions for the widths will be taken in a somewhat 
more general form~\cite{fn1}, mixing of states will be 
taken in the same simplified form as in Ref.~\cite{dpp}.
We further assume that only one nucleon-like resonance 
exists near 1700~MeV.

\subsection{Decays of $\Theta^+$}
\label{sec:theta}

We can write a partial width for the decay $B_1\to
M + B_2$ as \cite{fn1bis}
\be
\Gamma\left(B_1\to M B_2\right) = g^2_{B_1B_2M}
\cdot\frac{|\vec p|^3}{2\pi (M_1 + M_2)^2}\
\frac{M_2}{M_1} ,
\label{BWidth} \ee
where $|\vec{p}|$ is the c.m. momentum of the final
meson. In terms of the baryon masses $M_1, M_2$,
and the meson mass $m$, we have
\be
|\vec{p}| = \sqrt{[M_1^2 - (M_2 + m)^2]\cdot
[M_1^2 - (M_2 - m)^2]}/2M_1 \,.
\ee
Then, in the framework of the chiral soliton
approach, for the total width of $\Theta^+$ (summed
over two decay modes), we obtain~\cite{fn4}
\be
\Gamma\left(\Theta^+\to K N\right) = \frac{3}{5}\
\cdot\left(\cos\phi\cdot G_{\at} + \sin\phi\cdot
H_{\at} \frac{\sqrt 5}{4}\right)^2\cdot\frac{|\vec
p_{KN}|^3}{2\pi (M_\Theta + M_N)^2}\ \frac{M_N}
{M_\Theta} \,.
\label{ThetaW}\ee
Here, we admit the possibility of $SU(3)_{F}$-symmetry
breaking, which allows mixing of the baryon 
antidecuplet and octet states.  For simplicity, we mix 
here only a pair of soliton rotational states, the 
ground state $N$ and its rotational excitation 
$N^{\ast}$, which is a member of the same antidecuplet 
as $\Theta^+$.  As a result, instead of one coupling 
parameter $G_{\at}$, we obtain the set of three 
parameters $(H_{\at},~\sin\phi,~G_{\at})$.  The chiral 
soliton approach allows these to be determined, but 
with different levels of reliability. 

The constants $H_{\at}$ and $G_{\at}$ can be expressed 
in terms of universal constants $G_0, G_1$, and $G_2$. 
The corresponding expressions are (see \cite{kimpra,ekp}
for $H_{\at}$ and \cite{dpp} for $G_{\at}$):
\be
H_{\at}&=& G_0-\frac 52 G_1 +\frac 12 G_2\, ,\\
G_{\at}&=& G_0- G_1 -\frac 12 G_2\,.
\ee
Analysis of Ref.~\cite{dpp} showed that the constant
$G_2$ is very small, and in what follows we neglect 
it.  The octet coupling $G_8 = G_0 + G_1/2$ can be 
reliably extracted from properties of octet and 
decuplet baryons; in accordance with Ref.~\cite{dpp}, 
we take it here as $G_8\approx 18$.  In this way, we 
can relate constants $H_{\at}$ and $G_{\at}$ by 
$$2 G_{\at}-H_{\at}\approx G_8 \approx 18.$$  Keeping 
this in mind, we give below only values of $G_{\at}$.  
Note that at small $G_2$ and $G_1$, $G_{\at}\approx 
H_{\at}\approx G_8$, while at small $G_2$ and $G_{\at}$, 
$H_{\at}\approx -G_8$.

The mixing angle $\phi$ is less reliable.  Basing on 
its estimates in Ref.~\cite{dpp}, we use $\sin\phi
\approx 0.085\,$ (\textit{i.e.}, $\phi\approx
5^\circ$).

The coupling $G_{\at}$ is the least known quantity.
In the soliton picture, it receives different
contributions which tend to cancel each other,
making a definite conclusion difficult.  Nevertheless,
$G_{\at}$ was demonstrated to be suppressed, and the
conservative upper bound of $G_{\at}< 9.5$ has been
given~\cite{dpp}; moreover, it was shown that in the
non-relativistic limit for the quarks, the constant
$G_{\at}$ tends to zero (see also Ref.~\cite{mpr},
where it was shown that this cancellation happens for
any number of colors).  All these results suggest the 
coupling constant to be considerably smaller than its
upper limit~\cite{fn2}.

With expression (\ref{ThetaW}) and $M_{\Theta}=
1540$~MeV, the restrictions $\Gamma(\Theta^+\to
KN)\leq 1~[3;~5]~{\rm MeV}$ lead to $-1.4~[-2.9;~
{-4.0}]\leq G_{\at}\leq 2.9~[4.5;~5.6]\,$.  We do not
insist on the limit $\Gamma_{\Theta^+}\sim 1$~MeV as
given by Refs.~\cite{asw,asw1,cahntr}, as it is based
mainly on older experiments.  But we consider values
higher than 5~MeV to be improbable.

\subsection{Decays of $N^{\ast}$}
\label{sec:Nstar}

A large numerical value of $G_8$ (and $H_{\at}$), 
compared to $G_{\at}$, makes even a small 
octet--antidecuplet mixing a very important effect in 
decays of the $\Theta^+$.  It is even more important 
for decays of the $N^{\ast}$, the non-strange 
$P_{11}$ member of the antidecuplet (assumed in 
Ref.~\cite{dpp} to be identified with $N(1710)$).

First of all, the octet--antidecuplet mixing allows
the decay $N^{\ast}\to\pi\Delta$, otherwise forbidden 
for the pure antidecuplet member.  In addition, mixing 
essentially influences the partial decay width 
$N^{\ast}\to K\Lambda$.  A description of mixing 
effects in $N^{\ast}$ decays is most simple just for 
these two modes, since for them only the initial 
state $N^{\ast}$ has octet partner(s) to mix with. 
(Both the $\Delta$ and $\Lambda$ could mix with the 
antidecuplet members only under isospin violation.  In 
decays of the $\Theta^+$, only the final nucleon can 
mix, by assumption, just with $N^{\ast}$.)  These 
decay modes allow us to draw interesting conclusions
concerning the $N^{\ast}$.

The corresponding partial widths of the $N^{\ast}$
are \cite{fn6}:
\be
\Gamma(N^{\ast}\to \pi\Delta) &=& \frac{12}{5}\cdot
(\sin\phi\cdot G_{8})^2\cdot\frac{|\vec
p_{\pi\Delta}|^3}{2\pi (M_{N^\ast} +
M_\Delta)^2}\ \frac{M_\Delta^2}{M_{N^\ast}^2}\ , \\
\label{PWdel}
\Gamma(N^{\ast}\to K \Lambda) &=& \frac{3}{20}\cdot
\left(\cos\phi\cdot G_{\at} + \sin\phi\cdot G_8
\frac{4}{\sqrt 5}\right)^2\cdot\frac{|\vec
p_{K\Lambda}|^3}{2\pi(M_{N^\ast} + M_\Lambda)^2}\
\frac{M_\Lambda}{M_{N^\ast}}\ .
\label{PWlam} \ee
The decay widths here are summed over possible
charge states of the final hadrons.

Note that $\Gamma_{\ast}^{\pi\Delta}\equiv\Gamma(N
^{\ast}\to\pi\Delta)$ is independent of the very
uncertain antidecuplet coupling $G_{\at}$.  Using
the above-given values of other parameters, for
$M_{N^\ast}= 1680~[1730]~{\rm MeV}$, leads to
\be
\Gamma_{\ast}^{\pi\Delta}\approx 2.8~[3.5]~
{\rm MeV}\,
\label{delnum}\ee
and (for positive $G_{\at}$) $\Gamma_{\ast}
^{K\Lambda}\geq 0.17~[0.36]~{\rm MeV}\,$.  Taking
$G_{\at} = 2.9$ (the highest value compatible with
$\Gamma_{\Theta} = 1$~MeV) gives
\be
\Gamma_{\ast}^{K\Lambda}\approx 0.70~[1.56]~
{\rm MeV}\,,
\label{lamnum}\ee
again, for $M_{N^\ast} = 1680~[1730]$~MeV.

To investigate decays $N^{\ast}\to\pi N$, $\eta N$,
and $K\Sigma$, one needs to account for mixing of
both initial and final baryons.  Taking for coupling
constants the linear approximation in $\sin\phi$, we 
obtain
\be \label{PWpin}
\Gamma\left(N^{\ast}\to \pi N\right) &=& \frac{3}
{20}\cdot\left(G_{\at} - \frac{\sin\phi}{\sqrt 5}
\left[7 G_8-\frac{5}{4} H_{\at}\right]
\right)^2\cdot\frac{|\vec
p_{\pi N}|^3}{2\pi (M_{N^\ast} + M_N)^2} \
\frac{M_N}{M_{N^\ast}}\ ,\\
\label{PWeta}
\Gamma\left(N^{\ast}\to\eta N\right) &=&
\frac{3}{20}\cdot\left(G_{\at}
+ \frac{\sin\phi}{\sqrt 5}
\left[G_8-\frac{5}{4} H_{\at}\right]\right)^2
\cdot\frac{|\vec
p_{\eta N}|^3}{2\pi (M_{N^\ast} + M_N)^2}\
\frac{M_N}{M_{N^\ast}}\ ,\\
\label{PWsig}
\Gamma\left(N^{\ast}\to K\Sigma\right) &=&
\frac{3}{20}\cdot\left(G_{\at}
+ \frac{\sin\phi}{\sqrt 5}
\left[2 G_8+\frac{5}{2} H_{\at}\right]\right)^2
\cdot \frac{|\vec
p_{K\Sigma}|^3} {2\pi (M_{N^\ast} + M_\Sigma)^2}\
\frac{M_\Sigma}{M_{N^\ast}}\,.
\ee
For positive $G_{\at}$, it gives a cancellation
of the vertex for the $\pi N$ decay mode and 
enhancement due to mixing for the $\eta N$ mode 
(note in addition that the decay $N^{\ast}(1680)\to 
K\Sigma$ is forbidden and $N^{\ast}(1730)\to 
K\Sigma$ is suppressed by kinematics).  This picture 
does not look contradictory.  However, a detailed
description of the decays by expressions~
(\ref{PWpin}), (\ref{PWeta}), and (\ref{PWsig})
may be too simplified.  Indeed, Eq.~(\ref{PWpin})
with $G_{\at}= 2.9$ (which leads to the strongest
cancellation compatible with $\Gamma_{\Theta^+}=
1$~MeV) gives $$\Gamma_{\ast}^{\pi N} = 2.1~[2.3]
~{\rm MeV}$$ for $M_{N^\ast} = 1680~[1730]$~MeV,
in contradiction with the restrictions
(\ref{15}).  In other words, in such a simple
picture of mixing between only two nucleon-like
states, restrictions for $\Gamma_{\ast}^{\pi N}$
are incompatible with $\Gamma_{\Theta^+}\sim
1$~MeV.  However, a very small $\pi N$ partial
width $\Gamma_{\ast}^{\pi N}$ could be easily
accommodated by the soliton picture of the
octet--antidecuplet mixing if $\Gamma_{\Theta^+}
\sim 5$~MeV.

The situation can be changed if we take into
account mixing with one more nucleon-like state.
Such a case was suggested, \textit{e.g.}, by Jaffe
and Wilczek~\cite{JW}, as mixing with $N(1440)$.
However, they assumed the complete flavor
separation for a colored quark-antiquark pair,
which contradicts the existing understanding of
how the OZI-rule works.  We prefer a more
phenomenological version~\cite{dp}.  If the sign
of the mixing angle $\theta_N$, as introduced in
Ref.~\cite{dp}, is opposite to the sign of the
octet--antidecuplet mixing angle $\phi$, then the
additional contribution may provide additional
cancellation, and diminish the partial width of
the $\pi N$ decay mode, making it compatible with
$\Gamma_{\Theta^+}\sim 1$~MeV.  Regretfully, at
present, we can not determine the correct
relative sign for mixings of $N^{\ast}$ with $N$
and, say, $N(1440)$.  This would require a
knowledge of the nature and/or inner structure
of all the involved states.  Given the small
value of $G_{\at}$, we should not expect a large
mixing angle $\theta_N$ of $N^{\ast}$ with
$N(1440)$, as it would lead to rather large $\pi
N$ partial width of $N^{\ast}$.

Thus, in the framework of the soliton picture
(with mixing to additional nucleon-like states),
a $\Theta^+$-width of about 1~MeV implies that its
non-strange partner in the antidecuplet may have
a partial width of about 4~MeV for the
$SU(3)_F$-violating $\pi\Delta$ decay mode, up to
1~MeV for the $K\Lambda$ mode, and a couple of MeV 
for the $\eta N$ channel.  Decay to $K\Sigma$ 
should be small, if possible at all.  The total 
$N^{\ast}$-width, with all decay modes together, 
might achieve $\sim 10$~MeV, so that the $N^{\ast}$ 
state would be wider than the $\Theta^+$, though 
still narrow.

As a resonance in $\pi N$ collisions, the state
$N^{\ast}$ should be rather narrow and highly
inelastic, with a preference to decay mainly into
$\pi\pi N$ final states. Restrictions (\ref{15})
for $\Gamma_{\ast}^{\pi N}$ lead to a very small
elastic branching for $N^{\ast}\to\pi N$, not more
than 5\%.  Such a peak can not be extracted by
standard methods of PWA.

The assumption of a larger $\Gamma_{\Theta^+}$ (up
to about 5~MeV) does not influence the most intensive
decay channel $N^{\ast}\to\pi\Delta$.  Therefore, it
does not change the essential features of the above
conclusions.  Note that measurements of the ratio of 
$\pi N$ and $\eta N$ partial widths may provide us 
with valuable information about octet--antidecuplet 
mixing.

These findings together explain why the suggested
$N^{\ast}$ has not been observed up to now.  At the
same time, the above estimated width of the
$N^{\ast}$, as compared to $\Theta^+$, looks much
more reasonable for the antidecuplet member state
than the range of values tabulated for the $N(1710)$
by the PDG~\cite{PDG02}.

\subsection{Decays of $\Xi_{3/2}$}
\label{sec:Xi}

The above approach can be applied also to other
antidecuplet members.  We will not discuss here
decays of the $\Sigma$-like partner of the
$\Theta^+$, which should be more essentially
influenced by mixing(s).  Instead, we consider 
possible decays of the $\Xi$-like partner.

The quantum numbers and mass 1862~MeV of the $\Xi
_{3/2}$ admit 2-body decay modes $\pi\Xi$,
$\overline{K}\Sigma$, and $\pi\Xi(1530)$.
The latter decay would be analogous to $N^{\ast}\to
\pi\Delta$, but is stronger forbidden by the 
$SU(3)_F$-symmetry, since $\Xi_{3/2}$, contrary to
$N^{\ast}$, can not be mixed with octet members,
because of its different isospin.  The decay could
be allowed, nevertheless, by mixing of the
decuplet $\Xi(1530)$ with some octet
$\Xi$'s, which appears to be negligible.  Another
possibility for this decay would involve the
mixing of the $\Xi_{3/2}$ with a similar state
belonging to a higher $SU(3)_F$-multiplet, such
as the 27- or 35-plet.  However, at present, we
have not any definite information on such states.
That is why we do not discuss here the decay to
$\Xi(1530)$, though its experimental study
could give interesting and useful information.

For the partial widths of decay modes $\pi\Xi$
and $\overline K\Sigma$, we obtain expressions
similar to Eq.~(\ref{BWidth}):
\be
\Gamma\left(\Xi_{3/2}\to\pi\Xi\right) &=&
\frac{3}{10}\
\cdot
G_{\at}^2\cdot\frac{|\vec p_{\pi\Xi}|^3}{2\pi
(M_{\Xi_{3/2}} + M_{\Xi})^2}\ \frac{M_{\Xi}}
{M_{\Xi_{3/2}}} \,,
\label{Xipi} \\
\Gamma\left(\Xi_{3/2}\to\overline{K}\Sigma\right)
&=& \frac{3}{10}\ \cdot\left(\cos\phi\cdot
G_{\at} - \sin\phi\cdot H_{\at} \frac{\sqrt 5}{4}
\right)^2\cdot\frac{|\vec p_{\overline{K}
\Sigma}|^3}{2\pi (M_{\Xi_{3/2}} + M_{\Sigma})
^2}\ \frac{M_{\Sigma}}{M_{\Xi_{3/2}}} \,.
\label{antiKSig}
\ee
Note that the decay $\Xi_{3/2}\to\pi\Xi$ depends
only on $G_{\at}$.

For positive $G_{\at}$, consistent with
$\Gamma_{\Theta^+}\approx1$~MeV ({\textit i.e.},
$G_{\at}\approx 2.9$), these widths are the order
of a few MeV,
$$\Gamma\left(\Xi_{3/2}\to\pi\Xi\right)\approx
2.6~{\rm MeV}\,,~~~~~\Gamma\left(\Xi_{3/2}\to
\overline{K}\Sigma\right)\approx 2.0~{\rm MeV}\,.$$
On the other hand, for negative $G_{\at}$, again 
consistent with $\Gamma_{\Theta^+}\approx 1$~MeV 
(this time, $G_{\at}\approx -1.4$; note that 
negative values of $G_{\at}$ seem to be allowed 
theoretically), we obtain an extremely 
narrow $\Xi_{3/2}$

$$\Gamma\left(\Xi_{3/2}\to\pi\Xi\right)
\approx 0.6~{\rm MeV}\,,~~~~~\Gamma\left(\Xi_{3/2}\to
\overline{K}\Sigma\right)\approx 23~{\rm keV}\,,$$
with $\Gamma_{\Xi_{3/2}}$ perhaps somewhat
smaller than $\Gamma_{\Theta^+}$.  In both cases
$\Gamma_{\Xi_{3/2}}$ is small and would be very
difficult to measure directly.  Instead, there is
another interesting possibility.  The relative
intensity of the two $\Xi_{3/2}$-decay modes,
$\pi\Xi$ and $\overline{K}\Sigma$, may be very
different, depending on the manner of
$SU(3)_F$-violating mixing.  The branching ratio
for the latter mode may be either negligible, or
of the same order (or even larger) as compared to
the former one.  A measurement of this ratio
should be experimentally feasible.

\section{Conclusion and discussion}
\label{sec:disc}

To summarize, given our current knowledge of the
$\Theta^+$, the state commonly known as the $N(1710)$
is not the appropriate candidate to be a member of
the antidecuplet together with the $\Theta^+$. Instead,
we suggest candidates with nearby masses, $N(1680)$
(more promising) and/or $N(1730)$ (less promising, but
not excluded).  Our analysis suggests that the
appropriate state should be rather narrow and very
inelastic.  Similar considerations have been applied
to the $\Xi_{3/2}(1862)$, assumed to be also a member
of the same antidecuplet.  It should also be quite
narrow.

One can ask how definite are our theoretical
predictions.  They have, indeed, essential theoretical
uncertainties.  For example, the mixing angle $\phi$,
taken from Ref.~\cite{dpp}, was actually determined
through formulas containing the $\sigma$-term (just as
the mass difference in the antidecuplet). If we use
parameters corresponding to more recent information, 
for both the $\sigma$-term and the mass difference, we 
obtain larger mixing, up to $\sin\phi\approx 0.15$.  
With our formulas, this would most strongly influence 
the partial width $N^{\ast}\to\pi\Delta$, increasing 
it to about 15~MeV.  Other partial widths of $N^{\ast}$ 
change not so dramatically, and the total width 
appears to remain not higher than $\sim 30$~MeV.  Such 
a width could well be measured, but not in elastic 
scattering, because of an expected very small elastic 
branching ratio.  Note, however, that the above large 
value for $\sin\phi$ may appear problematic, since the 
formulas of Ref.~\cite{dpp} assume linearisation with 
respect to $SU(3)_F$-violation, and need to be 
reconsidered if the violation appears to be large.

A high degree of uncertainty emerges also because our 
approach can not definitely establish the existence of 
the resonance(s).  We have assumed the presence of only 
one state with $J^P = 1/2^+$, either $N(1680)$ or 
$N(1730)$.  If both exist, they should essentially mix 
due to their nearby masses, strongly changing our 
estimates.

Nevertheless, even having in mind all theoretical
uncertainties, we can suggest several directions for
experimental studies.  First of all, one should
search for possible new narrow nucleon state(s) in the
mass region near 1680 and/or 1730~MeV.  Searches may
use various initial states, (\textit{e.g.}, $\pi N$
collision or photoproduction).  We expect the largest
effect in the $\pi\pi N$ final state (though
$\pi\Delta$ is forbidden by $SU(3)_F$).  The final
states $\eta N$ and $K\Lambda$ may also be interesting
and useful, especially the ratio of $\eta N$ and $\pi
N$ partial widths as the latter is very sensitive to
the structure of the octet--antidecuplet mixing.  
Another interesting possibility to separate 
antidecuplet and octet components of $N^\ast$ is 
provided by comparison of photoexcitation amplitudes 
for neutral and charged isocomponents of this 
resonance, the point being that the antidecuplet 
component does not contribute to the photoexcitation 
of the charged component of $N^{\ast}$ (see details 
in Ref.~\cite{Polyakov:2003dx}).

For $\Xi_{3/2}$, attempts to measure the total width
are necessary,  though it could possibly be even
smaller than $\Gamma_{\Theta^+}$.  Branching ratios
for $\overline K\Sigma$ and $\pi\Xi(1530)$, in relation 
to $\pi\Xi$, are very interesting.  These may give 
important information on the mixing of antidecuplet 
baryons with octets and higher $SU(3)_F$-multiplets.

\acknowledgments

The authors thank D.~I.~Diakonov, V.~Y.~Petrov, and
P.~V.~Pobylitsa for many helpful discussions.
Special thanks are due to M.~Praszalowicz for
very valuable discussions.  Ya.~A. thanks Prof.~K.~
Goeke for hospitality extended to him in the
Institute for Theoretical Physics~II of the
Ruhr-University Bochum at final stages of the work.
The work was partly supported by the U.~S.~Department
of Energy Grant DE--FG02--99ER41110, by the Jefferson
Laboratory, by the Southeastern Universities Research
Association under DOE Contract DE--AC05--84ER40150,
by the Russian State Grant SS--1124.2003.2, and by
the Sofja Kovalevskaja Programme of the Alexander von
Humboldt Foundation, the German Federal Ministry of
Education and Research, and the Programme for
Investment in the Future of the German Government.


\newpage
\begin{table}[th]
\caption{\label{tbl1}
         Comparison of $\Theta^+$ and N(1710) properties.}
\begin{tabular}{cccc|cccc}
\colrule
Collaboration  & Mass            & Width     & Ref &
Collaboration  & Mass            & Width     & Ref \\
               & (MeV)           & (MeV)     &     &
               & (MeV)           & (MeV)     &     \\
\colrule
DPP            & 1530            & $<$15     &\protect\cite{dpp}   &
DPP            & 1710            &           &\protect\cite{dpp}   \\
\colrule
LEPS           & 1540$\pm$10     & $<$25     &\protect\cite{leps}  &&&&\\
DIANA          & 1539$\pm$2      & $<$9      &\protect\cite{itep1} &&&&\\
CLAS/$\gamma n$& 1542$\pm$5      & $<$21     &\protect\cite{clas1} &&&&\\
CLAS/$\gamma p$& 1540$\pm$10     & $<$32     &\protect\cite{clas2} &&&&\\
ELSA           & 1540$\pm$4$\pm$2& $<$25     &\protect\cite{elsa}  &&&&\\
ITEP/$\nu$     & 1533$\pm$5      & $<$20     &\protect\cite{itep2} &&&&\\
HERMES         & 1526$\pm$2.6$\pm$2.1& $<$19 &\protect\cite{herm}  &&&&\\
CLAS/$\gamma p$& 1555$\pm$10     & $<$26     &\protect\cite{clas3} &&&&\\
ZEUS           & 1527$\pm$2      & $<$24     &\protect\cite{zeus}  &&&&\\
NA49           & 1535            &           &\protect\cite{na49}  &&&&\\
\colrule
USC            & 1543            & $<$6      &\protect\cite{nuss}  &
KH             & 1723$\pm$9      &120$\pm$15 &\protect\cite{kh}    \\
GWU            & 1540$-$1550     & $\leq$1   &\protect\cite{asw}   &
CMU            & 1700$\pm$50     & 90$\pm$30 &\protect\cite{cut}   \\
J\"ulich       & 1545            & $<$5      &\protect\cite{hk}    &
KSU            & 1717$\pm$28     &480$\pm$230&\protect\cite{man}   \\
LBNL           & 1540            &0.9$\pm$0.3&\protect\cite{cahntr}&&&& \\
\colrule
\end{tabular}
\end{table}
\newpage
\newpage


\begin{thebibliography}{99}
\bibitem{fn1}  For example, it was assumed in~\cite{dpp} that
               parametrically $G_8/G_{\at}\sim 1$, on basis
               of large $N_c$ counting.  Here, we use a more
               phenomenological approach leaving $G_{\at}$
               unconstrained.
\bibitem{fn1bis}The factor $\frac{M_2}{M_1}$ is related to
               the nonrelativistic normalization of baryonic
               states in the soliton picture.  Note also
               that degree of this factor depends on the spin 
               of the participating baryons.
\bibitem{fn4}  We take this opportunity to correct a misprint
               in the corresponding expression of Ref.~
               \cite{dpp}.  The coefficient in front of its
               second term should be $\sqrt 5/2$ instead of
               $\sqrt 5/4$.
\bibitem{fn2}  Analysis of Rathke~\cite{Rathke}, performed
               for $G_{\at}$ in the framework of the chiral
               quark-soliton model~\cite{ChQSM}, gave a
               value of about 5, but it could be even
               smaller.
\end{thebibliography}
\end{document}